\documentclass[10pt,conference]{IEEEtran}
\IEEEoverridecommandlockouts
\usepackage{cite}
\usepackage{comment}
\usepackage{amsmath,amssymb,amsfonts}
\usepackage{algorithmic}
\usepackage{graphicx}
\usepackage{textcomp}
\usepackage{xcolor}
\def\BibTeX{{\rm B\kern-.05em{\sc i\kern-.025em b}\kern-.08em
    T\kern-.1667em\lower.7ex\hbox{E}\kern-.125emX}}
    
\begin{document}

\title{Static Analysis for Android GDPR Compliance Assurance\\}
\author{\IEEEauthorblockN{Mugdha Khedkar}
\IEEEauthorblockA{\textit{Heinz Nixdorf Institute} \\
\textit{Paderborn University} \\
Paderborn, Germany \\
mugdha.khedkar@upb.de}
}

\maketitle

\begin{abstract}
Many Android applications collect data from users. When they do, they must protect this collected data according to the current legal frameworks. Such data protection has become even more important since the European Union rolled out the General Data Protection Regulation (GDPR). App developers have limited tool support to reason about data protection throughout their app development process. Although many Android applications state a privacy policy, privacy policy compliance checks are currently manual, expensive, and prone to error. One of the major challenges in privacy audits is the significant gap between legal privacy statements (in English text) and technical measures that Android apps use to protect their user’s privacy.
In this thesis, we will explore to what extent we can use static analysis to answer important questions regarding data protection. Our main goal is to design a tool based approach that aids app developers and auditors in ensuring data protection in Android applications, based on automated static program analysis.
\end{abstract}


\section{Introduction}
We use several Android applications in our daily life, many of which collect data from us. While laws like the GDPR~\cite{1} exist to protect our data, these are frequently ignored by apps. Article 4 of the GDPR defines personal data as \textit{``any information relating to an identified or identifiable natural person, a data subject"} where an identifiable natural person is one who can be identified, directly or indirectly, by reference to an identifier such as a name, an identification number, etc. The GDPR imposes several obligations on the access, storage and processing of personal data. 

The growing demand for privacy by design\cite{2}, both by end users and by GDPR necessitates that app developers use state-of-the art technical measures to protect their users’ privacy. GDPR explicitly refers to one such technical measure for data protection, called \textit{pseudonymisation}. Pseudonymisation aims at protecting personal data by replacing parts of it with \textit{pseudonyms}. Personal data can then no longer be attributed to a specific data subject without the use of pseudonyms. The legal description of GDPR is very technical and long and hence it can be difficult for app developers to understand. 

Although many Android applications state a privacy policy, privacy policy compliance checks are currently manual. Violations of privacy policies can easily happen also inadvertently, in the worst case just by making use of data hungry third-party libraries. Since manual examination is expensive, auditors may be forced to sample randomly from the code under a source code audit. This can make the process unsound and error-prone.

Google Play recently launched a new feature, the data safety section~\cite{14} where app developers will be required to share how apps collect, share and secure users’ data. A recent study by Mozilla~\cite{15} has revealed discrepancies between the information reported in data safety sections and privacy policies of Android apps. Reporting privacy relevant information using these features needs manual effort from app developers. If we can reduce this manual effort required to develop and even audit Android apps that collect personal data from users, the entire process is likely to become much simpler, more accurate and less expensive. To automate this process, tools that bridge the legal and technical aspects of data protection are required. 

\section{Problem Statement}
We identify the following problems for Android app development and privacy auditing process:
\begin{enumerate}
\item Lack of tool support for app developers to write privacy-aware code proactively, as opposed to reactively.
\item A significant gap between legal privacy statements (in English text) and technical measures that Android apps use to protect their user’s privacy.
\item Lack of tool support for source code privacy audits.
\end{enumerate}
 In this work, we will explore if static analysis can be used to solve these problems. Static analysis checks the source code thoroughly before execution, and covers all of the app's possible execution paths. Thus the use of static analysis has the potential to eventually yield legally useful guarantees, and can be effectively used to aid app developers and auditors alike. 

\section{Research Questions}    
We will answer the following research questions as part of this thesis: 

\textbf{RQ1.} To what extent can precise static analysis aid privacy-aware Android app development and auditing? 

\textbf{RQ2.} Which static analysis methods can be used to diagnose and explain data protection issues? Are current methods sufficient, or should we design new static analysis techniques?

\textbf{RQ3.} How can we bridge the gap between legal privacy statements and the technical measures Android apps use for data protection? 

\section{Related Work}

Much of the existing work on static taint analysis~\cite{3,4,5,6} is aimed at finding security vulnerabilities (by tracking taint flows from sources to sinks). MUDFLOW\cite{7} studies data flows from sensitive sources to sensitive sinks in both benign and malicious Android apps. The most interesting observation is that most accesses to sensitive data do not end up in sensitive sinks. This means that existing taint analysis tools will not report them. Hence there is a need to perform an exploratory analysis on sources of sensitive data. Our focus is to explore how personal data is being processed through an Android app, irrespective of whether it causes a data leak.

Furthermore, most taint analysis tools~\cite{3,4,5,7} use a fixed set of sources and sinks as a starting point for taint analysis. In our work, we will focus only on data that can be categorised as personal data with respect to GDPR. Hence we need a different mechanism to label personal data. Existing tool-based approaches~\cite{8,9} can identify sensitive user-input data in Android applications but do not categorise data as personal data. Unfortunately, these are in-house proprietary tools and hence cannot be used in our work.

Jicer~\cite{11} is a static program slicer that works with an intermediate representation of Java code. We intend to use Jicer to slice the application code to understand and represent how personal data flows through the code.

Recent work by Feiyang Tang and Bjarte M. Østvold~\cite{10} describes an automatic software analysis technique that characterises the flow of privacy-related data. The results of the analysis can be presented as a graph of privacy flows and operations, and this representation can be understood by software developers and auditors alike. Their work differs from our proposed work in two ways. First, they use a predetermined list of sources and sinks but do not categorise data as personal data. Secondly, it is aimed at Data Protection Impact Management~\cite{12} and answers only limited questions. 

\section{Expected Contributions}
\label{contributions}
The expected contributions of this thesis are as follows:
\begin{enumerate} 
\item To design a tool-based approach based on static program analysis that aids app developers in writing privacy-aware code throughout the app development process.
\item To use this tool for assisting privacy audits, reducing their costs by a large margin.
\end{enumerate}

\section{Solution Approach}
Our work will focus on Android apps. While apps also process data on server backends, this code is not usually available to outsiders. Thus, we will first concentrate on the client side of the apps, while also making explicit the assumptions about the backend processing on the server side.  We plan to achieve our research goals through the following modules:

    \subsection{Labelling sources of personal data in Android apps}
    To perform any static analysis on potential personal data, one requires a reliable mechanism to detect and label the sources of personal data that a given Android app processes. Some of this data, called system centric private data, is provided by the OS for eg. through system calls like \textit{getLastKnownLocation()}. However, one also needs to label the data that the user provides as input to an Android application, called user-provided private data. We have manually constructed a dataset of personal data keywords and are currently developing a tool that identifies the sources of user-provided personal data and labels them to ensure the soundness of our analysis. We will refine this dataset periodically and explore whether using NLP techniques can ensure the correctness of our tool.
    
    \subsection{Designing a suitable data representation}
    \label{dr}
    Once we have identified the sources of personal data, we will slice the program to understand and represent how this data flows through the source code. Slicing will be based on data and control flow of labelled personal data, and the resulting slices will be represented in a way that is understandable by both technical and legal experts. To this end, we will extend Jicer~\cite{11} to examine and visualise the resulting privacy-relevant program slices to aid app developers and auditors. This data representation will express source code components in terms of legal aspects of GDPR using a data protection vocabulary~\cite{13}. We will use this data representation to perform a risk analysis of the program slices and highlight parts of the application code which process raw personal data. As a consequence, we will already be able to give some hints to the developers and the highlighted program slices will also help auditors examine relevant code fragments instead of randomly sampling the code.

    \subsection{Building a comprehensive prototype}
    Next we will use the data representation we design in \ref{dr} to answer questions regarding data protection. When paired with an appropriate static analysis tool, it will  help answer important questions for a program slice: \textit{Is personal data pseudonymised along all paths? What data is derived from the collected personal data, which third parties is it being sent to? Is personal data being sent to Analytics engines? How is personal data being manipulated?} Designing such a static analysis will simplify the automated and tool-assisted reasoning about the privacy-compliance of Android applications, further enabling privacy by design. 
    
    \subsection{Evaluation on open source Android apps}
    We will evaluate our prototype on open source Android apps that are known to process personal data, such as Telegram, K-9 mail, Keepass2, Signal etc. We will conduct in-depth case studies and generate a report containing evidence of the technical measures taken by these apps to protect personal data. We will also conduct several user studies to get feedback from app developers and auditors. This module will enable deployment of our research artefacts into the industry and assist in assuring source code compliance.

\end{document}